\documentclass[aps,prd,preprint,superscriptaddress,tightenlines,%
nofootinbib]{revtex4}
\newcommand{\PRE}[1]{{#1}}   

\usepackage{bm}
\usepackage{epsfig}

\newcommand{\postscript}[2]{\setlength{\epsfxsize}{#2\hsize}
   \centerline{\epsfbox{#1}}}

\newcommand{\md}{M_D}
\newcommand{\mbh}{M_{\text{BH}}}
\newcommand{\mbhmin}{M_{\text{BH}}^{\text{min}}}

\newcommand{\xmin}{x_{\text{min}}}
\newcommand{\ifb}{\text{fb}^{-1}}
\newcommand{\ev}{\text{eV}}

\newcommand{\gev}{\text{GeV}}
\newcommand{\tev}{\text{TeV}}
\newcommand{\pb}{\text{pb}}
\newcommand{\cm}{\text{cm}}
\newcommand{\km}{\text{km}}
\newcommand{\g}{\text{g}}
\newcommand{\s}{\text{s}}
\newcommand{\ns}{\text{ns}}
\newcommand{\yr}{\text{yr}}
\newcommand{\sr}{\text{sr}}

\newcommand{\xmax}{X_{\text{max}}}
\newcommand{\etal}{{\em et al.}}

\newcommand{\eqref}[1]{Eq.~(\ref{#1})}
\newcommand{\Enu}{E_{\nu}}

\begin{document}

\preprint{
\hfil
\begin{minipage}[t]{3in}
\begin{flushright}
\vspace*{.4in}
NUB--3229--Th--02\\
UCI--TR--2002--22\\
UK/02--09\\
hep-ph/0207139
\end{flushright}
\end{minipage}
}

\title{
\PRE{\vspace*{1.5in}}
Neutrino Bounds on Astrophysical Sources and New Physics
\PRE{\vspace*{0.3in}}
}

\author{Luis A.~Anchordoqui}
\affiliation{Department of Physics,\\
Northeastern University, Boston, MA 02115
\PRE{\vspace*{.1in}}
}

\author{Jonathan L.~Feng}
\affiliation{Department of Physics and Astronomy,\\
University of California, Irvine, CA 92697
\PRE{\vspace*{.1in}}
}

\author{Haim Goldberg}
\affiliation{Department of Physics,\\
Northeastern University, Boston, MA 02115
\PRE{\vspace*{.1in}}
}

\author{Alfred D.~Shapere}%
\affiliation{Department of Physics,\\
University of Kentucky, Lexington, KY 40506
\PRE{\vspace*{.5in}}
}

\begin{abstract}
\PRE{\vspace*{.1in}}
Ultra-high energy cosmic neutrinos are incisive probes of both
astrophysical sources and new TeV-scale physics.  Such neutrinos would
create extensive air showers deep in the atmosphere. The absence of
such showers implies upper limits on incoming neutrino fluxes and
cross sections. Combining the exposures of AGASA, the largest existing
ground array, with the exposure of the Fly's Eye fluorescence detector
integrated over all its operating epochs, we derive 95\% CL bounds
that substantially improve existing limits.  We begin with
model-independent bounds on astrophysical fluxes, assuming standard
model cross sections, and model-independent bounds on new physics
cross sections, assuming a conservative cosmogenic flux. We then
derive model-dependent constraints on new components of neutrino flux
for several assumed power spectra, and we update bounds on the
fundamental Planck scale $M_D$ in extra dimension scenarios from black
hole production. For large numbers of extra dimensions, we find $M_D >
2.0\ (1.1)~\tev$ for $\mbhmin = M_D\ (5M_D)$, comparable to or
exceeding the most stringent constraints to date.
\end{abstract}

\pacs{96.40.Tv, 13.15.+g, 04.50.+h, 04.70.-s}

\maketitle

\section{Introduction}

Cosmic neutrinos provide a unique window on astrophysical processes
because they escape from dense regions and typically propagate to the
Earth unhindered~\cite{Sigl:2001th}.  At ultra-high energies, they
also provide an important probe of new ideas in particle physics.  In
contrast to all other standard model (SM) particles, their known
interactions are so weak that new physics may easily alter neutrino
properties, sometimes drastically.  This is especially relevant for
neutrinos with energies above $10^6~\gev$, which interact with
nucleons with center-of-mass energies above 1 TeV, where the SM is
expected to be modified by new physics.

The signal for ultra-high energy neutrinos is quasi-horizontal giant
air showers initiated deep in the
atmosphere~\cite{Anchordoqui:2002hs}.  This signal is well-studied and
easily differentiated from air showers initiated by hadrons.  The
Earth's atmospheric depth rises from about $1000~\g/\cm^2$ vertically
to nearly $36000~\g/\cm^2$ horizontally.  For all but the most extreme
(and typically problematic~\cite{Burdman:1997yg,Kachelriess:2000cb})
neutrino cross sections, the mean free path of neutrinos is larger
than even the horizontal atmospheric depth. Neutrinos therefore
interact with roughly equal probability at any point in the atmosphere
and may initiate showers in the volume of air immediately above the
detector. These will appear as ``normal'' showers, with large
electromagnetic components, curved fronts (a curvature radius of a few
km), and signals well spread over time (of the order of microseconds).

On the other hand, hadrons have interaction lengths of the order of
$40~\g/\cm^2$ and so always interact at the top of the atmosphere.
The electromagnetic component of an air shower has mean interaction
length $\sim 45-60~\g/\cm^2$.  For a quasi-horizontal shower initiated
by an ordinary hadron, then, this component is absorbed long before
reaching the ground, as it has passed through the equivalent of
several vertical atmospheres --- 2 at a zenith angle of $60^\circ$, 3
at $70^\circ$, and 6 at $80^\circ$. In these showers, only high energy
muons created in the first few generations of particles survive past 2
equivalent vertical atmospheres. The shape of the resulting shower
front is therefore very flat (with curvature radius above $100~\km$),
and its time extension is very short (less than $50~\ns$).

These shower characteristics are exploited by both ground arrays and
fluorescence detectors in searches for primary cosmic ray neutrinos.
At present, no ultra-high energy neutrino signal has been reported.
Here we determine the total exposure for neutrino detection from
existing facilities and derive both model-independent and
model-dependent bounds on astrophysical neutrino fluxes and new
neutrino interactions.

The outline of the paper is as follows. In Sec.~\ref{sec:exposure} we
examine acceptances for neutrino detection and compute the current
combined total exposure using all available data from the Akeno Giant
Air Shower Array (AGASA)~\cite{Chiba:1991nf} and Fly's
Eye~\cite{Baltrusaitis:mx} experiments.  In Sec.~\ref{sec:fluxes} we
determine model-independent bounds on the total neutrino flux,
assuming SM cross sections. To derive model-independent results, we
assume only that fluxes are confined to a small window around some
central neutrino energy and obtain bounds as a function of this
central energy. After that, we assume a power law neutrino flux
$d\Phi/dE_{\nu} \propto E_\nu^{-\gamma}$ to obtain stronger, but more
model-dependent, bounds on the total neutrino flux from integrating
over all energies.

In Sec.~\ref{sec:interactions} we derive model-independent bounds on
high energy neutrino cross sections, assuming a conservative
cosmogenic flux.  These significantly improve existing
limits~\cite{Tyler:2001gt}. We then derive model-dependent bounds on
cross sections, focusing on the example of TeV-scale gravity scenarios
in Sec.~\ref{sec:gravity}.  We improve existing
constraints~\cite{Anchordoqui:2001cg} on the fundamental Planck scale
from the non-observation of microscopic black hole production by
cosmic neutrinos~\cite{Feng:2001ib,Anchordoqui:2001ei,%
Emparan:2001kf,Ringwald:2001vk,Kowalski:2002gb}. For large numbers of
extra dimensions, these bounds are comparable to or exceed all
existing bounds on extra dimensions. Sec.~\ref{sec:conclusions}
contains our conclusions.

\section{Neutrino exposure}
\label{sec:exposure}

To estimate the sensitivity of a ground-based detector to
neutrino-initiated showers, we must first compute its exposure for
various types of showers.  The exposure is the product of the
effective aperture and the range of depths within which the shower
must originate to trigger the device, integrated over time.  The
effective aperture is the detector's projected area weighted by
detection probability and integrated over solid angle.  The exposure
is, then,
\begin{equation}
{\cal E} (E_{\text{sh}}) \approx \int_0^T\, dt\, \int_0^{h_{\rm max}} \,
(A\Omega)_{\rm eff}(E_{\text{sh}},t) \,
\frac{\rho_0}{\rho_{\rm water}} \, e^{-h/H}\, dh \, , \label{b}
\end{equation}
where $T$ is the total observation time of the detector, $h_{\rm max}
= 15~\km$, $H \approx 8~\km$, and $\rho_0 \approx 1.15 \times 10^{-3}
\rho_{\rm water}$ is the density of the atmosphere at ground level.
The effective aperture is~\cite{Billoir:nq}
\begin{equation}
(A\Omega)_{\rm eff} (E_{\text{sh}},t) \equiv \int_{\theta_{\rm
min}}^{\theta_{\rm max}} A(t)\, {\cal P}(E_{\text{sh}},\theta,t)\,
2\pi\, \sin\theta \, d\theta
\end{equation}
where $A(t)$ is the detector's area, ${\cal P}
(E_{\text{sh}},\theta,t)$ is the probability that a shower that
arrives with energy $E_{\text{sh}}$ and zenith angle $\theta$ triggers
the detector, and the angular cuts $\theta_{\rm min}$ and $\theta_{\rm
max}$ are chosen to optimize detection efficiency while eliminating
hadronic background.

Exposures at a given detector depend on detection method and shower
type.  We will refer to showers as hadronic or electromagnetic (EM),
depending on the nature of their first interaction, irrespective of
their later development. With this convention, for example, the SM
neutral current process $\nu N \to \nu X$ produces a hadronic shower,
while the charged current process for {\em electron} neutrinos $\nu_e
N \to e X$ produces both a hadronic shower and an EM shower.

\begin{figure}
\postscript{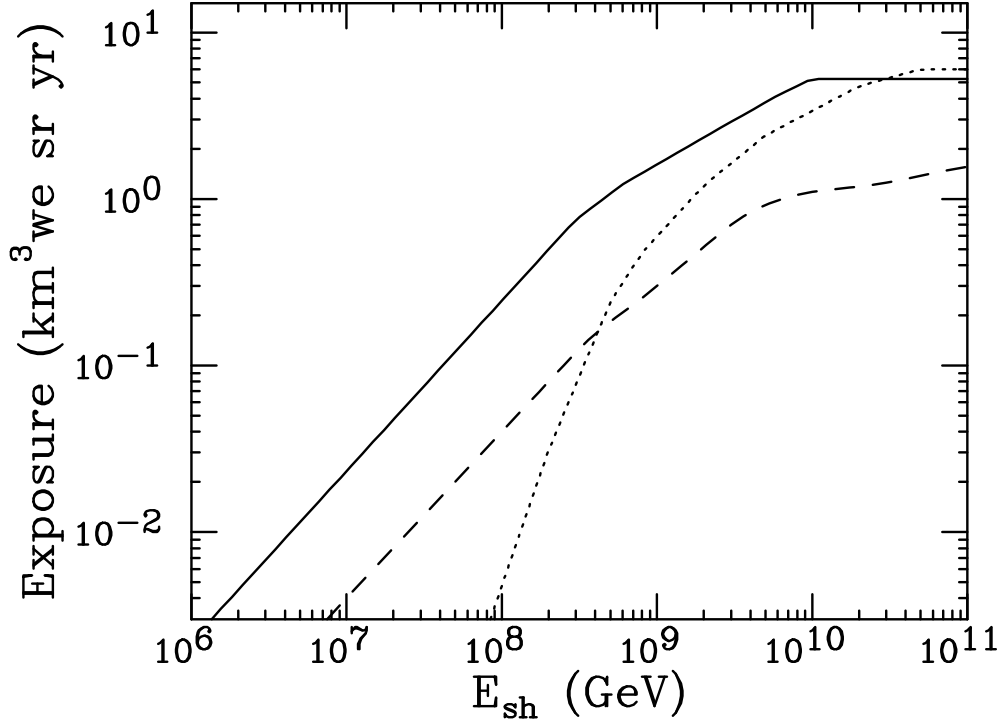}{0.80}
\caption{The figure shows the total exposures for hadronic showers at
AGASA (solid), EM showers at AGASA (dashed), and EM and hadronic
showers at Fly's Eye (dotted) as functions of shower energy.}
\label{exposure}
\end{figure}

For ground arrays, exposures for hadronic and EM showers
differ~\cite{Capelle:1998zz}. In hadronic showers, the initial
hadronic interaction produces a strong muon component that remains
until the shower reaches the ground.  This muon component is largely
absent for EM showers.  These muons significantly enhance triggering
efficiencies for ground arrays, which are sensitive only to ground
level activity, and so exposures for hadronic showers exceed exposures
for EM showers at ground arrays. Above some critical energy, which
depends on the total effective area of the array, the detector
exposure saturates. Exposures for hadronic and EM showers for the
AGASA ground array~\cite{Chiba:1991nf} are given in
Fig.~\ref{exposure}. These exposures are for $1.5 \times 10^{8}~\s$ of
livetime between December 1995 and November 2000. The hadronic
exposure was derived in Ref.~\cite{Anchordoqui:2001cg}, based on
results from searches for deeply penetrating showers~\cite{Inoue:cn}
and conservative assumptions. We derive the EM exposure by comparison
with results for the Auger experiment.  Effective apertures for Auger
have been calculated in Ref.~\cite{Capelle:1998zz}.  For AGASA, we
adopt the Auger aperture for quasi-horizontal EM showers with axes
falling in the array, reduced by a factor of 30, the ratio of surface
areas of the arrays.  We have checked that, using this EM exposure, we
reproduce AGASA's bounds~\cite{Yoshida} on $\nu_e$ fluxes to within
20\%.

In contrast to ground arrays, the hadronic and EM exposures of
fluorescence detectors are very similar. Fluorescence detectors are
sensitive to the total EM activity along the entire longitudinal
development of the shower.  In EM showers, essentially all of the
energy produces EM activity.  In hadronic showers, hadronic collisions
produce equal numbers of $\pi^0$, $\pi^+$, and $\pi^-$.  The $\pi^0$
decays to photons, producing EM energy.  However, the charged pions
typically interact before decaying.  Through successive interactions,
most of their energy also becomes EM activity.  As a result, roughly
90\% of the energy in hadronic showers is EM.  The response and
efficiency of fluorescence detectors to hadronic and EM showers is
therefore expected to be similar~\cite{Baltrusaitis:mt}. In this work
we adopt the total Fly's Eye exposure reported in
Ref.~\cite{Baltrusaitis:mx} for both hadronic and EM showers. This
exposure, given in Fig.~\ref{exposure}, includes not only data from
the first epoch of observation (February 1983 to May 1985), most of
which were reported in Ref.~\cite{Baltrusaitis:mt}, but also data from
four additional running periods between November 1985 to July 1992.
The additional periods enhance the total Fly's Eye exposure by roughly
a factor of 3.

The AGASA Collaboration has searched for quasi-horizontal showers that
are deeply-penetrating, with depth at shower maximum $X_{\rm max} >
2500$~g/cm$^2$~\cite{Yoshida}. At AGASA, the location of the shower
maximum is determined through its correlation to two measurable
quantities:~$\eta$, which parameterizes the lateral distribution of
charged particles at ground level, and $\delta$, which parameterizes
the curvature of the shower front. Deeply penetrating events must
satisfy $\xmax^{\eta}, \xmax^{\delta} \ge 2500~\g/\cm^2$.  The
expected background from hadronic showers is
$1.72{}^{+0.14}_{-0.07}{}^{+0.65}_{-0.41}$, where the first
uncertainty is from Monte Carlo statistics, and the second is
systematic.  Among the 6 candidate events, 5 have values of
$\xmax^{\eta}$ and/or $\xmax^{\delta}$ that barely exceed
$2500~\g/\cm^2$, and are well within $\Delta \xmax$ of this value,
where $\Delta \xmax$ is the estimated precision with which $\xmax$ can
be reconstructed. The AGASA Collaboration thus concludes that there is
no significant enhancement of deeply penetrating shower rates given
the detector's resolution.

The Fly's Eye observes an air shower as a nitrogen fluorescence light
source traveling at the speed of light through the
atmosphere~\cite{Baltrusaitis:mx}. The light is emitted isotropically
with intensity proportional to the number of charged particles in the
shower. The received light profile is reconstructed by a 3 parameter
fit to the charged particle density (largely electrons and positrons)
along the shower track. The parameters are the depth of the observed
first interaction $X_0$; the depth of the shower maximum $X_{\rm
max}$; and the density $N_{\rm max}$ at $X_{\rm max}$.  In a running
time of eleven years, Fly's Eye recorded more than 5000
events. However, there are no neutrino candidates with shower maximum
deeper than
$2500~\g/\cm^2$~\cite{Baltrusaitis:mx,Baltrusaitis:mt,Gaisser:ix}.

All in all, given 1 event that unambiguously passes all cuts with 1.72
events expected from hadronic background, the combined results of
AGASA and Fly's Eye imply an upper bound of 3.5 events at 95\%
CL~\cite{Feldman:1997qc} from neutrino fluxes.

The event rate for quasi-horizontal deep showers from ultra-high
energy neutrinos is
\begin{equation}
N = \sum_{i,X} \int dE_i\, N_A \, \frac{d\Phi_i}{dE_i} \, \sigma_{i
N \to X} (E_i) \, {\cal E}_{iX}(E_i)\ , \label{numevents}
\end{equation}
where the sum is over all neutrino species $i = \nu_e, \bar{\nu}_e,
\nu_{\mu}, \bar{\nu}_{\mu}, \nu_{\tau}, \bar{\nu}_{\tau}$, and all
final states $X$. $N_A = 6.022 \times 10^{23}$ is Avogadro's number,
and $d\Phi_i/dE_i$ is the source flux of neutrino species $i$.
Finally, ${\cal E}_{iX}(E_i)$ is the appropriate exposure measured in
$\cm^3\text{we}~\sr \cdot \text{time}$.

To clarify, we present appropriate exposures for some SM processes.
At the ultra-high energies of interest, on average 20\% of the total
neutrino energy goes into hadronic recoil for both SM neutral current
and charged current events~\cite{Gandhi:1995tf}. Exposures for SM
charged current events at AGASA are therefore
\begin{eqnarray}
{\cal E}_{\nu_e X}(E_{\nu_e}) &=&
\min \{ {\cal E}_{\text{had}}(0.2 E_{\nu_e}) +
{\cal E}_{\text{EM}}(0.8 E_{\nu_e}), {\cal E}_{\text{sat}} \} \\
{\cal E}_{\nu_{\mu} X}(E_{\nu_{\mu}}) &=&
{\cal E}_{\text{had}}(0.2 E_{\nu_{\mu}}) \\
{\cal E}_{\nu_{\tau} X}(E_{\nu_{\tau}}) &=& {\cal
E}_{\text{had}}(0.2 E_{\nu_{\tau}}) \ ,
\end{eqnarray}
with identical expressions for anti-neutrinos.  The exposures for
$\nu_{\mu}$ and $\nu_{\tau}$ are identical, because at these energies,
$\tau$ leptons typically do not decay before arriving at the Earth's
surface.  For $\nu_e$, the expression includes the effects of
saturation, noted previously.  For AGASA, the exposure saturates at
${\cal E}_{\text{sat}} \approx 5.3~\km^3\text{we}~\sr~\yr$ at
$E_{\text{sh}} \approx 10^{10}~\gev$ (see Fig.~\ref{exposure}).  For
neutral current interactions at AGASA, all of the right-hand entries
are ${\cal E}_{\text{had}}(0.2 E_{\nu_i})$.

For Fly's Eye, the corresponding charged current exposures are simply
\begin{eqnarray}
{\cal E}_{\nu_e X}(E_{\nu_e}) &=&
{\cal E}( E_{\nu_e})\\
{\cal E}_{\nu_{\mu} X}(E_{\nu_{\mu}}) &=&
{\cal E}(0.2 E_{\nu_{\mu}}) \\
{\cal E}_{\nu_{\tau} X}(E_{\nu_{\tau}}) &=& {\cal E}(0.2
E_{\nu_{\tau}}) \ .
\end{eqnarray}
For neutral current exposures at Fly's Eye, all of the right-hand
entries are ${\cal E}(0.2 E_{\nu_i})$.

Given the exposures ${\cal E}$ described above, the absence of events
implies upper bounds on cross sections, assuming some fixed flux, or
upper bounds on fluxes, assuming some fixed cross section.  We now
consider these two possibilities in turn.

\section{Bounds on astrophysical neutrino fluxes}
\label{sec:fluxes}

To derive bounds on the neutrino flux, we assume SM charged and
neutral current interactions for all neutrinos.  We further assume
that the source flux of neutrinos, which in the energy range of
interest is dominantly $\nu_{\mu}$, $\bar{\nu}_{\mu}$, and $\nu_e$ at
production, is completely mixed in flavor upon arrival at the Earth.
There is now strong evidence for maximal mixing among all neutrino
species~\cite{Fukuda:1998mi}.  In addition, given mass differences of
$\Delta m^2 \agt 10^{-6}~\ev^2$, the neutrino flux is expected to be
completely mixed if they travel a distance $\agt 0.1~\text{Mpc}$.  The
assumption of equal flavor representation upon arrival is therefore
strongly supported by data~\cite{Athar:2000yw}.

We first derive model-independent bounds on the total neutrino flux.
Let us start by noting that if the number of events integrated over
energy is bounded by 3.5, then it is certainly true bin by bin in
energy. Thus, using Eq.~(\ref{numevents}) we obtain
\begin{equation}
\sum_{i,X} \int_{\Delta} dE_i\, N_A \, \frac{d\Phi_i}{dE_i} \,
\sigma_{i N \to X} (E_i) \, {\cal E}_{iX}(E_i)\  < 3.5\ ,
\label{bound}
\end{equation}
at 95\% CL for some interval $\Delta$. Here the sum over $X$ takes
into account charge and neutral current processes.  In a logarithmic
interval $\Delta$ where a single power law approximation
\begin{equation}
\frac{d\Phi_i}{dE_i} \, \sigma_{i N \to X} (E_i) \, {\cal E}_{iX}(E_i)
\sim E_i^{\alpha}
\end{equation}
is valid, a straightforward calculation shows that
\begin{equation}
\int_{\langle E\rangle e^{-\Delta/2}}^{\langle E\rangle e^{\Delta/2}}
\frac{dE_i}{E_i} \,
E_i\, \frac{d\Phi_i}{dE_i} \, \sigma_{i N \to X} (E_i) \, {\cal
E}_{iX}(E_i)  =  \langle \sigma_{\nu_i N\rightarrow X} (E_i)\,
{\cal E}_{iX}(E_i)\,  E_i\,
d\Phi_i/dE_i \rangle
\frac{\sinh \delta}{\delta}\, \Delta\ ,
\label{sinsh}
\end{equation}
where $\delta=(\alpha+1)\Delta/2$ and $\langle A \rangle$ denotes the
quantity $A$ evaluated at the center of the logarithmic interval.  The
parameter $\alpha = 0.363 + \beta - \gamma$, where the 0.363 is the
power law index of the SM neutrino cross
sections~\cite{Gandhi:1995tf}, and $\beta$ and $-\gamma$ are the power
law indices (in the interval $\Delta$) of the exposure and flux
$d\Phi_i/dE_i$, respectively.  Since $\sinh \delta/\delta >1$, a
conservative bound may be obtained from Eqs.~(\ref{bound}) and
(\ref{sinsh}):
\begin{equation}
N_A\, \sum_{i,X} \langle \sigma_{\nu_i N\rightarrow X} (E_i)
\rangle \, \langle {\cal E}_{iX}(E_i)\rangle\, \langle E_i
d\Phi_i/dE_i \rangle < 3.5/\Delta\ . \label{avg}
\end{equation}
In this work we choose $\Delta=1$, corresponding to one $e$-folding of
energy, as a likely interval in which the single power law behavior is
valid. By setting $\langle E_i d\Phi_i/dE_i \rangle = \frac{1}{6}
\langle \Enu d\Phi_{\nu}/d\Enu\rangle$, where $\Phi_{\nu}$ is the
total neutrino flux, we obtain model-independent upper limits on the
total neutrino flux at 95\% CL. The results are given in
Table~\ref{table}.

\begin{table}
\caption{Model-independent upper limits on the differential neutrino
flux at 95\% CL.}
\begin{tabular}{ccc} \hline \hline
$E_\nu$ (GeV) & \hspace{7.5cm}
&$ \langle \Enu d\Phi_{\nu}/d\Enu \rangle$
(km$^{-2}$ sr$^{-1}$ yr$^{-1})$ \\
\hline $1\times 10^{8}$ & & $1.8 \times 10^{5}$ \\
$3\times 10^8$ & & $4.1\times 10^4$\\
$1\times 10^{9}$ & & $7.9 \times 10^{3}$ \\
$3\times 10^9$ & & $2.2\times 10^3$ \\
$1\times 10^{10}$ & &  $5.0 \times 10^{2}$ \\
$3\times 10^{10}$ & & $1.6 \times 10^2$ \\
$1\times 10^{11}$ & & $6.8 \times 10^{1}$ \\
\hline \hline \label{table}
\end{tabular}
\end{table}

\begin{figure}
\postscript{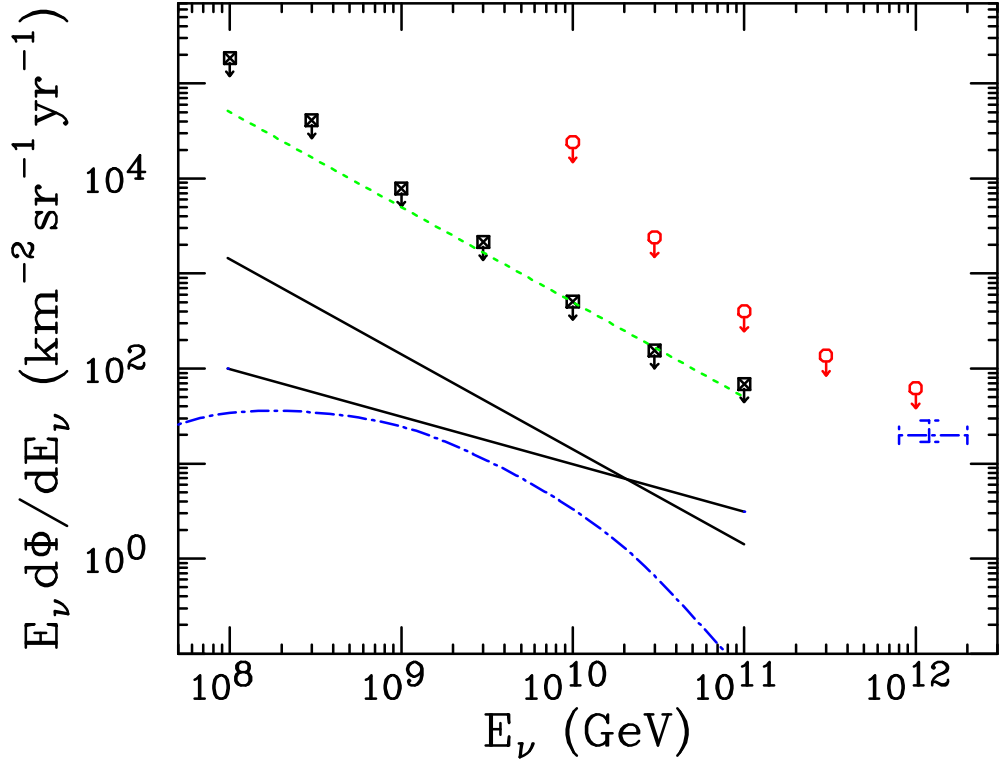}{0.80}
\caption{95\% CL model-independent (squares) and model-dependent
(solid lines) upper limits on the total neutrino flux $E_\nu d\Phi_\nu
/dE_\nu$ derived here using the combined exposures of AGASA and Fly's
Eye.  The model-dependent bounds assume $d \Phi_{\nu} / d E_{\nu}
\propto E^{-\gamma}$, with $\gamma = 1.5$ and 2.  For comparison, we
also present limits on the $\nu_\mu + \nu_e$ from GLUE
(circles)~\cite{Gorham:2001aj} and the 95\% CL bound from RICE on the
$\nu_e$ flux assuming a power law spectrum with $\gamma=2$
(dashed)~\cite{Kravchenko:2002mm}. The dash-dotted line is the
predicted total cosmogenic flux from pion
photo-production~\cite{Protheroe:1996ft}, and the point with error
bars is the total neutrino flux required by $Z$-burst models, as
derived in~\cite{Fodor:2002hy}.  }
\label{fluxes}
\end{figure}

These model-independent upper bounds on the total neutrino flux can be
strengthened by assuming a particular flux behavior. For example, if
the neutrino flux falls like
\begin{equation}
\frac{d\Phi_\nu}{dE_\nu} = J_0
\left(\frac{E_\nu}{E_0}\right)^{-\gamma} \ ,
\end{equation}
Eq.~(\ref{bound}) leads to $J_0 < 1.4 \times
10^{-5}~\km^{-2}~\sr^{-1}~\yr^{-1}~\gev^{-1}$, for $\gamma = 2$ and
$E_0 = 10^8~\gev$. Under the same assumptions for $\gamma = 1.5$ one
obtains $J_0 < 9.8 \times
10^{-7}~\km^{-2}~\sr^{-1}~\yr^{-1}~\gev^{-1}$. Fig.~\ref{fluxes} shows
both the model-independent bounds of Table~\ref{table} as well as the
bounds on flux under the power law assumptions just
discussed. Additionally displayed in the figure are the upper limits
on the $\nu_{\mu}+\nu_e$ flux obtained from the non-observation of
microwave \v{C}erenkov pulses from EM showers induced by neutrinos in
the Moon's rim, as measured by the Goldstone Lunar Ultra-high energy
neutrino Experiment (GLUE)~\cite{Gorham:2001aj}, as well as bounds
from the search for radio pulses from EM showers created by electron
neutrino collisions in ice at the Radio Ice \v{C}erenkov Experiment
(RICE)~\cite{Kravchenko:2002mm}. Comparing model-independent bounds to
model-independent bounds, and model-dependent bounds to
model-dependent bounds, we find that the bounds obtained here are
significantly more stringent than existing limits.

Also displayed in Fig.~\ref{fluxes} as a single point with error
bars is the total neutrino flux required by
$Z$-bursts~\cite{Weiler:1982qy}. The $Z$-burst model proposes
that the observed ultra-high energy cosmic rays with energies
above $10^{11}~\gev$ are secondaries resulting from resonant
annihilation of ultra-high energy neutrinos on relic neutrinos at
the $Z$ pole~\cite{Fargion:1997ft}.  A recent
analysis~\cite{Fodor:2002hy} includes several possibilities for a
diffuse background of protons, as distinct from protons resulting
from the $Z$-burst itself. The point shown in the figure
corresponds to the case in which the proton background originates
at distances $\alt 50$ Mpc. The normalized Hubble expansion rate,
the matter and vacuum energy densities, and the maximum redshift
are taken as $h=0.71$, $\Omega_M=0.3$, $\Omega_{\Lambda}=0.7$,
and $z_{\rm max}=2$, respectively.  The neutrino flux is assumed
to have no cosmological evolution. The horizontal errors result
from the $1\sigma$ uncertainty in the neutrino mass
determination. The errors in the flux reflect the statistical
fluctuations in the fits, as well as the uncertainty in the
Hubble expansion rate.

A more speculative explanation of the mysterious events at the high
end of the spectrum assumes that the cosmic ray primaries arise in the
decay of massive elementary $X$ particles. Sources of these exotic
particles could be either topological defects left over from early
universe phase transitions ($m_X \sim 10^{16} -
10^{19}~\gev$)~\cite{Hill:1982iq}, or some long-lived metastable
superheavy ($m_X \agt 10^{12}~\gev$) relic particles produced through
vacuum fluctuations during the inflationary stage of the
universe~\cite{Gondolo:1991rn}. The $X$ particles typically decay to
leptons and quarks. The latter produce jets of hadrons containing
mainly pions, together with a 3\% admixture of nucleons. The predicted
spectrum would thus be dominated by gamma rays and neutrinos produced
via pion decay. The neutrino flux bounds derived in this work
therefore seriously constrain this type of model. Moreover, recent
analyses of Haverah Park data~\cite{Ave:2000nd} suggest that less than
50\% of the primary cosmic rays above $4\times 10^{10}~\gev$ can be
photons at 95\% CL.  Note that mechanisms which successfully deplete
the high energy photons (such as efficient absorption on the universal
and galactic radio background) require an increase in the neutrino
flux to maintain the overall normalization of the observed
spectrum~\cite{Barbot:2002kh}. Definite quantitative comparison with
the neutrino flux bounds presented here can be obtained by specifying
the nature of the decay of the $X$ particle.

\section{Bounds on new physics interactions}
\label{sec:interactions}

In this section we examine the potential of probing the big desert
that lies between the electroweak and grand unified theory (GUT)
scales using ultra-high energy neutrino interactions. To derive bounds
on possible new physics contributions to neutrino cross sections, we
assume the ``guaranteed'' flux of cosmogenic neutrinos arising from
pion photo-production from ultra-high energy protons propagating
through the cosmic microwave background.  This flux depends on the
cosmological evolution of the cosmic ray sources. Throughout this
work, we conservatively adopt the estimates of Protheroe and
Johnson~\cite{Protheroe:1996ft} with nucleon source spectrum scaling
as $d\Phi_{_{{\rm nucleon}}} / dE \propto E^{-2}$ and extending up to
the cutoff energy $10^{12.5}~\gev$. We assume also a cosmological
source evolution scaling as $(1+z)^4$ for redshift
$z<1.9$~\cite{Engel:2001hd}. The total ultra-high energy cosmogenic
neutrino flux is shown in Fig.~\ref{fluxes}.

We consider this flux highly conservative.  For example, one might
conjecture that the cosmogenic flux is absent because the observed
cosmic rays with energies $\agt 10^{11}~\gev$ are protons or nuclei
generated by nearby sources within 50 Mpc.  However, none of the known
nearby candidate sources, such as Virgo, M82, Centaurus A, and
galactic pulsars~\cite{Ahn:1999jd} is as powerful as more distant
sources, such as Cygnus A and Pictor A~\cite{Carilli}.  The latter
must therefore inject nucleons with energies of at least
$10^{12}~\gev$. Indeed, contributions to the nucleon channel from
semi-local sources~\cite{Hill:1985mk}, as well as recently discussed
possibilities~\cite{Kalashev:2002kx} of source spectra harder than
$\Enu^{-1.5}$ and source evolutions stronger than $(1+z)^4$, would all
significantly enhance the cosmogenic neutrino flux.  Contributions
from decaying topological defects, active galactic nuclei, and other
speculative sources would have a similar effect.  If realized in
nature, any one of these possibilities would strengthen our bounds.

For SM interactions and the cosmogenic neutrino flux given in
Fig.~\ref{fluxes}, the expected rate for deeply penetrating showers at
AGASA and Fly's Eye is about 0.02 events per year, and so
negligible. New physics may contribute to neutrino cross sections in a
multitude of
ways~\cite{Feng:2001ib,Emparan:2001kf,Nussinov:1999jt,Domokos:1986qy},
with different contributions for different flavors, and a variety of
final states producing showers with a variety of hadronic and EM
shower components.  Here we assume that the new physics is
flavor-blind, inducing equivalent cross section modifications for all
neutrino species, and that the resulting new physics final state leads
to showers with negligible EM component. We then consider two simple
but representative cases: (1) $y = 1$, and (2) $y = 0.1$, where $y
\equiv E_{\text{sh}}/E_{\nu}$ is the average inelasticity.  In case
(1), the shower energy $E_{\text{sh}}$ differs little from the
neutrino energy $E_{\nu}$, as in TeV-scale black hole production (see
below). In case (2), $E_{\text{sh}}$ is substantially less than
$E_{\nu}$. This holds, for example, when neutral current interactions
are enhanced by the exchange of Kaluza-Klein (KK)
gravitons~\cite{Kachelriess:2000cb}.

\begin{figure}
\postscript{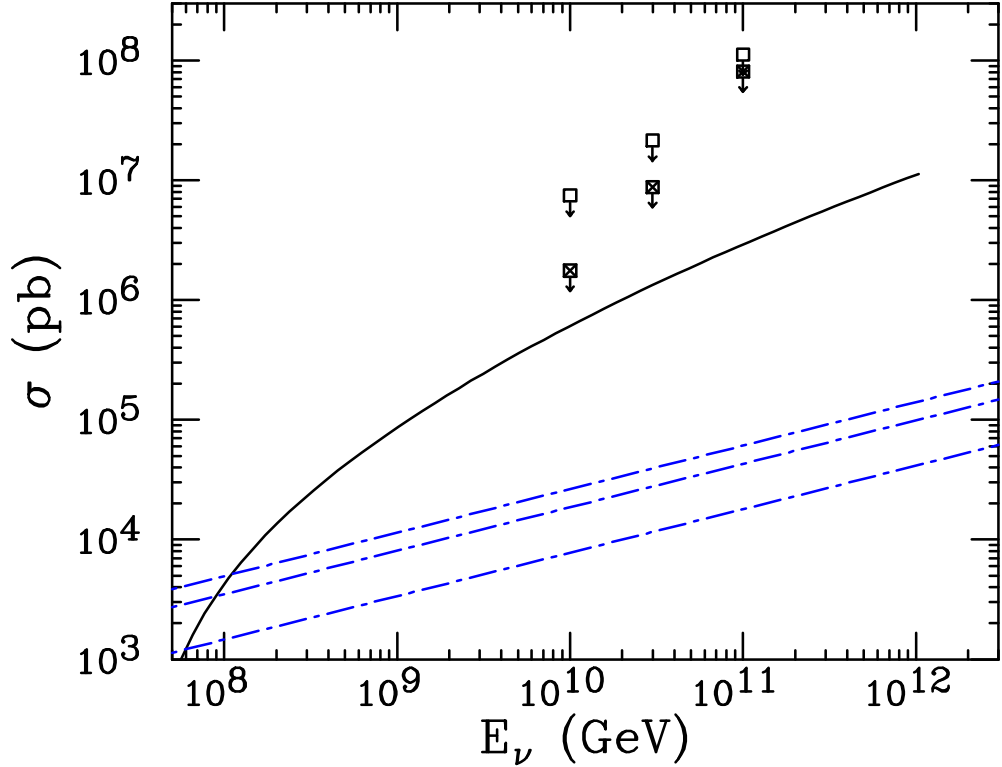}{0.80}
\caption{95\% CL model-independent upper limits on $\sigma_{\nu N}$
for inelasticity $y = 1$ (filled squares) and $y=0.1$ (open
squares). The solid contour is the upper limit on the black hole
production cross section for $n=7$ and $\xmin=5$, corresponding to
$M_D = 1.1~\tev$.  For comparison, the SM neutral current, charged
current, and total neutrino nucleon cross sections are also given
(dot-dashed, from below).  These bounds assume that the neutrino cross
sections does not exceed $6\times 10^8~\pb$ (see text).}
\label{sigmas}
\end{figure}

Our starting point to derive model-independent bounds on the total
neutrino-nucleon cross section $\sigma_{\nu N}$ is again
Eq.~(\ref{avg}).  For reasons given previously, we choose $\Delta =1,$
so that Eq.~(\ref{avg}) becomes
\begin{equation}
N_A \,  \langle \sigma_{\nu N \rightarrow X} (E_\nu) \rangle \,
\langle {\cal E} (y E_\nu) \rangle\, \langle E_{\nu}
d\Phi_\nu/dE_\nu\rangle < 3.5 \ . \label{csbound}
\end{equation}
Using the exposures in Fig.~\ref{exposure} and the cosmogenic flux in
Fig.~\ref{fluxes}, we find model-independent bounds on the neutrino
cross section for different energies and the two inelasticities
above. The resulting limits are given in Table~\ref{table2} and shown
in Fig.~\ref{sigmas}. For reference, the SM cross
sections~\cite{Gandhi:1995tf} are also given in the figure.

\begin{table}
\caption{95\% CL model-independent upper limits on the neutrino-nucleon
cross section.}
\begin{tabular}{ccc}
\hline
\hline
$E_\nu$ (GeV)
& \hspace{5cm} $\sigma_{\nu N}^{y = 1}$~(pb) \hspace{5cm}
& $\sigma_{\nu N}^{y = 0.1}$~(pb) \hspace{5cm}\\
\hline
$1\times 10^{10}$ & $1.8 \times 10^6$ &  $7.4 \times 10^6$ \\
$3\times 10^{10}$ & $8.8 \times 10^6$ & $2.2 \times 10^7$ \\
$1\times 10^{11}$ & $8.1 \times 10^7$ & $1.1 \times 10^8$ \\
\hline \hline \label{table2}
\end{tabular}
\end{table}

These bounds assume that neutrinos produce deeply penetrating
quasi-horizontal showers.  They may be avoided if neutrinos are so
strongly interacting that they shower high in the atmosphere.  For
zenith angles of $70^\circ$, this requires an interaction length
below $3000~\g/\cm^2$, corresponding to a cross section above $6
\times 10^8~\pb$. If the interaction length is between 3000
g/cm$^2$ and the horizontal atmospheric depth of 36000 g/cm$^2$,
corresponding to a $\nu N$ cross section between $6\times 10^8$~pb
and $5\times 10^7$~pb, respectively, an exponential decrease in
the event rate for showers of a given total energy will be
observed beyond a critical angle. Since this range includes our
upper bounds on $\sigma_{\nu N}$ at $10^{11}$ GeV, it is
conceivable that $\sigma_{\nu N}$ could thus be measured
directly.  Bounds on anomalous $\nu N$ cross sections which are
near-hadronic strength can be obtained through considerations of
their feedback to low energy physics via dispersion
relations~\cite{Goldberg:1998pv}.

The bounds shown in Table~\ref{table2} strengthen previous
bounds~\cite{Tyler:2001gt} by roughly one order of magnitude.
This enhancement results from a number of factors.  First, the
exposure used in Ref.~\cite{Tyler:2001gt} was limited to the
first $6 \times 10^6~\s$ of Fly's Eye operation. The updated
exposure used here is roughly 9 times larger.  Second, the
cosmogenic flux used in Ref.~\cite{Tyler:2001gt} assumes a source
cutoff energy of $10^{11.5}~\gev$, and is smaller by roughly a
factor of 4 than the one used here. We find it unnatural to
assume that the the source cutoff coincides with the maximum
observed cosmic ray energy, and so have used the flux
corresponding to the larger cutoff energy.  Finally, we have
presented 95\% CL limits, corresponding to a limit of 3.5 events.

As in the case of flux bounds, the model-independent cross section
bounds of Table~\ref{table2} will be improved if one assumes a cross
section shape and so can integrate over all energies.  We consider a
particular example in the next section.

\section{Implications for TeV-scale gravity}
\label{sec:gravity}

The idea that our universe could be a brane embedded in some higher
dimensional world has received a great deal of renewed attention over
the last 5 years~\cite{Antoniadis:1990ew,Randall:1999ee}. From a
phenomenological point of view, this possibility presents a new
perspective on the hierarchy between the gravitational and electroweak
mass scales. In these scenarios, the effective 4-dimensional Planck
scale $M_{\rm Pl} \sim 10^{19}~\gev$ is determined by the fundamental
$(4+n)$-dimensional Planck scale $M_D \sim 1~\tev$ and the geometry of
the $n$ extra dimensions.

Arguably the most fascinating prediction of TeV-scale gravity is the
production of black holes (BHs) in observable collisions of elementary
particles~\cite{Banks:1999gd,Cheung:2001ue}. For cosmic rays, this
implies that ultra-high energy neutrinos may produce BHs in the
atmosphere, initiating deep quasi-horizontal showers far above SM
rate~\cite{Feng:2001ib}. BH production therefore provides a specific
example of a model-dependent cross section that is bounded by the
arguments discussed above.

TeV-scale gravity also has a number of other implications for cosmic
rays.  The implications of perturbative KK graviton exchange has been
considered in a number of
scenarios~\cite{Nussinov:1999jt}.  However, for cosmic
rays, in contrast to the case at colliders, there is an abundance of
center-of-mass energy.  Extra dimensional effects will therefore first
appear as non-perturbative BH production in processes with
center-of-mass energies above the fundamental Planck scale, rather
than through perturbative effects below the Planck scale.  This is in
stark contrast to the case at colliders, where the sensitivity to KK
graviton effects surpasses the sensitivity to BH
production.\footnote{We thank S.~Dimopoulos for emphasizing this
point.}

The sensitivity of current cosmic ray experiments to BH production, as
well as that of facilities expected in the not-too-distant future, has
been thoroughly investigated~\cite{Anchordoqui:2001cg,%
Feng:2001ib,Emparan:2001kf,Ringwald:2001vk,Kowalski:2002gb}.
The parton-parton cross section is
estimated from the geometric area of the BH horizon and is of order
$\hat \sigma_i \sim \pi r_s^2$~\cite{Banks:1999gd}, where
\begin{equation}
\label{schwarz}
r_s(\mbh) =
\frac{1}{\md}
\left[ \frac{\mbh}{\md} \right]^{\frac{1}{1+n}}
\left[ \frac{2^n \pi^{(n-3)/2}\Gamma({n+3\over 2})}{n+2}
\right]^{\frac{1}{1+n}}
\end{equation}
is the radius of a Schwarzschild BH in $4{+}n$
dimensions~\cite{Myers:un}.  Criticisms of the absorptive black
disc scattering amplitude, which center on the exponential
suppression of transitions involving a (few-particle) quantum
state to a (many-particle) semiclassical
state~\cite{Voloshin:2001vs}, have been addressed in
Refs.~\cite{Dimopoulos:2001qe,Giddings:2001ih}.  The geometric
cross section applies  for both flat and
hyperbolic~\cite{Kaloper:2000jb} extra dimensions that are larger
than the Schwarzschild radius, and for warped extra dimensions
where $r_s$ is small compared to the curvature scale of the
geometry associated with the warped
subspace~\cite{Giddings:2000mu,Anchordoqui:2002fc}.

The total production cross section for BHs with mass $\mbh \equiv
\sqrt{sx}$ is then~\cite{Feng:2001ib}
\begin{equation}
\sigma_{\nu N \to \text{BH}} (\Enu) =
\sum_i \int_{(\mbhmin{})^2/s}^1 dx\,
\hat{\sigma}_i ( \sqrt{xs} ) \, f_i (x, Q) \ ,
\label{partonsigma}
\end{equation}
where $s = 2 m_N \Enu$, $x$ is the parton momentum fraction, the
$f_i$ are parton distribution functions (pdf's), $\mbhmin$ is the
minimum BH mass, and the sum is carried out over all partons in
the nucleon.  The choice of the momentum transfer $Q$ is
governed by considering the time or distance scale probed by the
interaction. According to Thorne's hoop
conjecture~\cite{Thorne:ji}, the formation of a well-defined
horizon in four dimensions occurs when the colliding particles
are at a distance $\sim r_s$ apart. (Note that there could be an
$n$-dependent factor for higher number of dimensions.) This has
led to the advocacy of the choice $Q\simeq
r_s^{-1}$~\cite{Emparan:2001kf}, which has the advantage of a
sensible limit at very high energies. However, as has been
pointed out by Dimopoulos and Emparan~\cite{Dimopoulos:2001qe},
string progenitors can give experimental  signals akin to BH for
the collision energies and values of $M_{\rm BH}^{\rm min}$ under
present consideration. In cosmic ray experiments, these signals
are indistinguishable from those of BH decay (more on this
below). Detailed calculations~\cite{Cheung:2002aq} show that
these ``string ball'' cross sections can exceed BH cross
sections. In this region, the dual resonance picture of string
theory would suggest a choice $Q\sim M_{\rm res}\sim \sqrt{xs}.$
Since we have chosen to use only BH cross sections over the
entire energy region, we set $Q = \min \{ \mbh, 10~\tev \}$,
where the upper limit is from the CTEQ5M1 distribution
functions~\cite{Lai:2000wy}. We are aware that for $Q\agt$ string
scale, the pdf's will receive significant corrections from the
rapid increase of degrees of freedom. Fortunately, as noted in
Ref.~\cite{Anchordoqui:2001cg}, the cross section $\sigma_{\nu N
\to {\rm BH}}$ is largely insensitive to the details of the choice
of $Q.$ For example, the two choices discussed here result in
cross sections that differ by only 10\% to 20\%. 

Once produced, BHs will Hawking evaporate with a temperature
proportional to the inverse radius $T_H=(n+1) / (4 \pi r_s)$. The
wavelength $\lambda = 2\pi/T_H$ corresponding to this temperature is
larger than the BH size. Hence, to first approximation the BH behaves
like a point-radiator with entropy
\begin{equation}
\label{entropy}
S = {4\pi \, \mbh\ r_s\over n+2}
\end{equation}
and mean lifetime
\begin{equation}
\tau_{_{\rm BH}} \sim {1\over \md} \left({\mbh\over \md}\right)^{3+n
\over 1+n}\ .
\end{equation}
Microscopic black holes therefore decay almost instantaneously to a
thermal distribution of SM particles. As very few SM particles are
invisible to cosmic ray detectors, the neutrino energy is almost
entirely transformed into shower energy.  The EM component of these
showers differs substantially from that of SM neutrino interactions,
allowing a good characterization of the phenomenon against background
when the BH entropy~$\gg$ 1~\cite{Anchordoqui:2001ei}. Recently, it
has been noted that BH recoil induced by KK-graviton emission may
launch the BH out of the brane~\cite{Frolov:2002as}.  In this case, BH
radiation would be prematurely terminated from the perspective of
brane observers.  This effect would drastically deplete the rate of
deeply developing showers, and could be misinterpreted as a sharp
cutoff on the ultra-high energy neutrino spectrum. We assume here that
the effect of recoil is negligible.

An important parameter in determining the BH cross section is $\xmin
\equiv \mbhmin / M_D$, the ratio of the minimal black hole mass to the
fundamental Planck scale.  The above description of BH production and
decay relies on semi-classical arguments, valid for large $\xmin$ or,
equivalently, large BH entropy.  For large $\xmin$, thermal
fluctuations due to particle emission are small ($S \gg
1$)~\cite{Preskill:1991tb}, statistical fluctuations in the
microcanonical ensemble are small ($\sqrt{S} \gg 1$), and quantum
gravity effects may be safely neglected.  In addition, gravitational
effects of the brane on BH production, which are ignored in all
analyses to date, are expected to be insignificant for BH masses well
above the brane tension, which is presumably $\sim M_D$. None of these
is true for $\xmin \approx 1$.

Cosmic ray experiments are largely insensitive to the exact details of
BH decay.  Whatever happens near $\xmin \approx 1$, it seems quite
reasonable to expect that BHs or their Planck mass progenitors will
decay visibly, triggering deeply atmospheric cascade
developments. There is, however, sensitivity to the BH production
cross section.  In string theory, BH production is expected to
gradually pass to the regime of string ball production as $\mbh$
approaches $M_D$.  Evidence from this picture suggests that the BH
production cross section is not radically altered in this
limit~\cite{Dimopoulos:2001qe}.  This addresses many of the concerns
listed above, but does not address the problem of brane effects on BH
production --- these may still be large for $\xmin \approx 1$.  In our
analysis, we avoid choosing a specific $\xmin$; rather, we present
results for the generous range $1 \leq x_{\rm min} \leq 10$. As we
will see, the bounds are rather insensitive to $\xmin$, in contrast
to the case at colliders such as the LHC.

\begin{figure}
\postscript{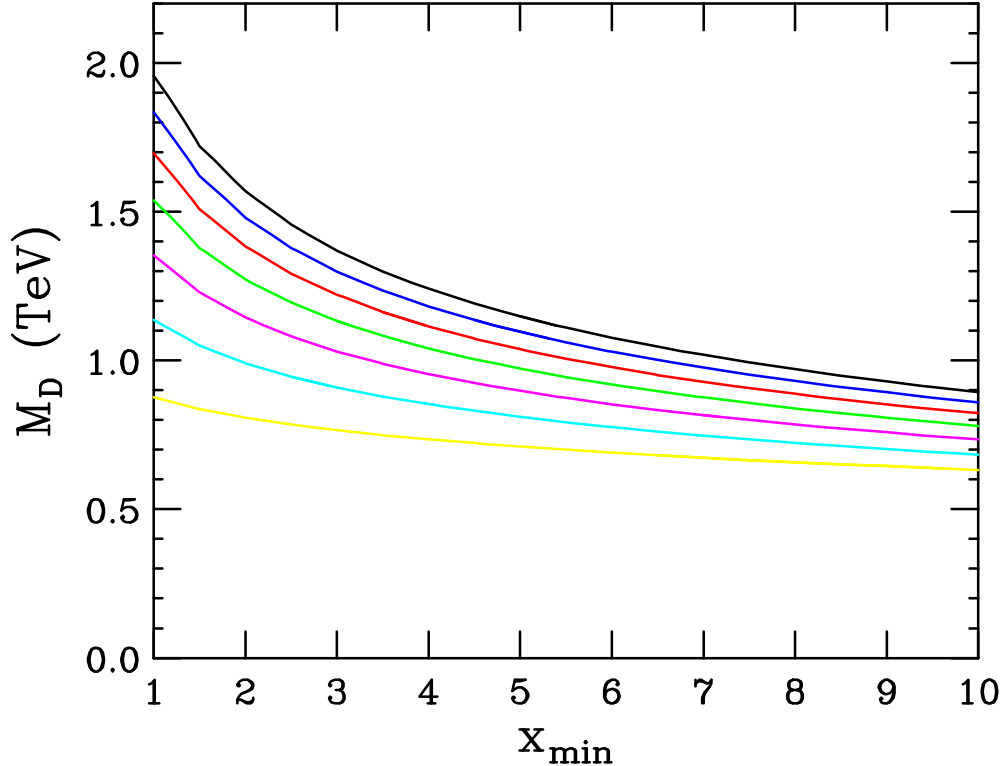}{0.80}
\caption{95\% CL lower limits on the fundamental Planck scale as a
function of $\xmin$ for $n = 1, \ldots, 7$ extra dimensions (from
below).}
\label{limits}
\end{figure}

In Fig.~\ref{limits} we show the lower limit on $\md$ as a function of
$\xmin$ corresponding to $<3.5$ events (95\% CL) observed for the
combined exposures of AGASA and Fly's Eye.  Bounds on $\md$ from
table-top gravity experiments, as well as from astrophysical and
cosmological considerations, greatly exceed 1 TeV, for models with $n
\leq 4$ flat extra dimensions~\cite{Hoyle:2000cv}. For $n > 4$,
however, the bounds of Fig.~\ref{limits} are among the most stringent
to date.  Note that these bounds do not depend on the shape of the
extra dimensions and are valid for both warped and non-warped
scenarios~\cite{Anchordoqui:2002fc}. For $\xmin = 1$, the bounds
extend up to 2.0 TeV for $n=7$.  Moreover, assuming $\xmin =3$, for
which the entropy $S> 10$, the bounds derived with the combined
exposure, for $n = 5, 6 ,7$, are $\md > 1.26~\tev, 1.30~\tev,
1.40~\tev$, respectively. All of these exceed bounds from the Tevatron
and LEP~\cite{Abbott:2000zb}, even in the case where the brane
softening parameter $\Lambda$~\cite{Murayama:2001av} is as large as
$\md$ (see~\cite{Anchordoqui:2001cg} for details).\footnote{The
increased exposure used here will also strengthen existing
bounds~\cite{Anchordoqui:2002it} from $p$-brane
production~\cite{Ahn:2002mj} in asymmetric compactifications.}

In Fig.~\ref{sigmas} we have also plotted the maximal BH cross
section, corresponding to $M_D = 1.1~\tev$, for the case $n=7$ and
$\xmin=5$.  As expected, given a model for the cross section's energy
dependence, the resulting bounds on new interactions are much more
stringent than the model-independent limits derived above.

\begin{figure}
\postscript{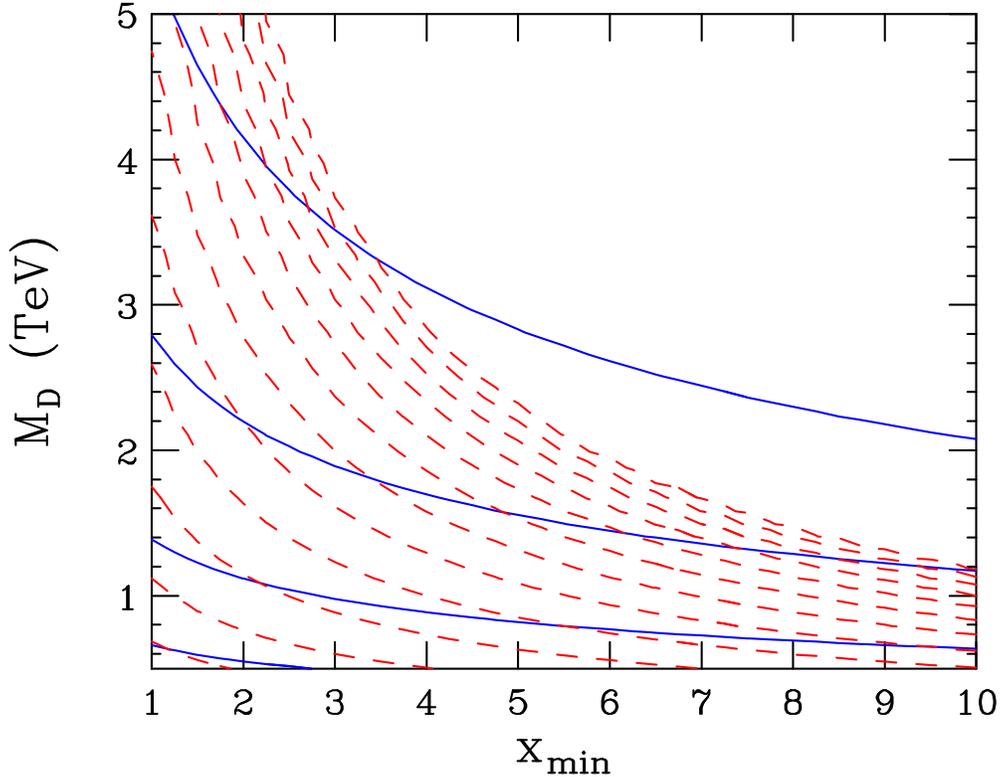}{0.80}
\caption{Black hole events at Auger in 3 years (1, 10, 100, 1000, from
above) (solid) and at the LHC for integrated luminosity $100~\ifb$ (1,
10, 100, $\ldots$ , $10^{11}$, from above) (dashed) for $n=7$ extra
dimensions.}
\label{fig:augerLHC_7}
\end{figure}

Finally, as noted above, the event rates for black hole
production by cosmic rays are fairly insensitive to the choice of
$\xmin$.  This contrasts sharply with the case at colliders. Specifically, 
the total production cross section of a BH of mass
$M_{\rm BH} \sim \sqrt{\tau s}$ in a $pp$ collision is given by
\begin{equation}
\sigma_{pp \to {\rm BH}} (\tau_{\rm min}, s) = \sum_{ij}
\int_{\tau_{\rm min}}^1 d \tau \int_\tau^1 \frac{dx}{x} f_i(x)
f_j(\tau/x) \hat\sigma_{ij}\,,
\end{equation}
where $\tau$ is the parton-parton center-of-mass energy squared
fraction, and $\sqrt{\tau_{\rm min} s}$ is the minimum
center-of-mass energy for which the black disc approximation is
valid. The number of BH produced at the LHC ($\sqrt{s} = 14$~TeV
and luminosity ${\cal L} =
10^{34}$~cm$^{-2}$~s$^{-1}$~\cite{Evans:1999dk}) is then $N^{\rm
LHC} = \int \sigma_{pp \to {\rm BH}}\, {\cal L}\, dt.$ In
Fig.~\ref{fig:augerLHC_7}, we show both Auger~\cite{Beatty:ff} and 
LHC event rates
for various $\xmin$.  For fixed $\md$, the LHC event rates drop
by one to two orders of magnitude for every unit increase in
$\xmin$, while the Auguer event rates are relatively stable. This
may be understood as resulting from a combination of the very
high energies available in cosmic neutrinos and the fact that the
parton energy is not degraded by a parton distribution function
in the cosmic neutrino `beam.'

\section{Conclusions}
\label{sec:conclusions}

In the first part of this paper, we derived new limits on the cosmic
neutrino flux striking the Earth's atmosphere. This was accomplished
by searching for quasi-horizontal deeply developing showers in
ultra-high energy cosmic ray data, taking into account the combined
exposures of the AGASA and Fly's Eye experiments. Our results
significantly strengthen existing limits and present serious problems
for models where exotic elementary $X$ particles cascade decay to
cosmic ray particles. In particular, models where topological defects
are responsible for the events detected with energies $\agt
10^{11}~\gev$ are severely constrained, because neutrinos are
typically a significant component in $X$ decays, and have a hard
spectrum extending up to $M_{\rm GUT} \sim 10^{16}~\gev$. The bounds
obtained in this paper will also challenge any attempt to normalize
the observed spectrum to the proton flux as predicted by top down
models.

In the second part of the paper, we used the atmosphere as a giant
calorimeter to probe neutrino-nucleon cross sections at $\sqrt{s} \agt
1~\tev$. We first combined the complete neutrino exposure of the
above-mentioned facilities with the flux of cosmogenic neutrinos, to
derive model-independent upper bounds on the neutrino-nucleon cross
section.  These bounds strengthen existing limits by roughly one order
of magnitude.  We then considered TeV-scale gravity models to study BH
production. The upper bounds on the neutrino-nucleon cross section
implied lower limits on the fundamental Planck scale, which represent
the best existing limits on TeV-scale gravity for $n \geq 5$ extra
spatial dimensions.

\begin{acknowledgments}
JLF thanks Savas Dimopoulos for conversations concerning black holes.
The work of LAA and HG has been partially supported by the US National
Science Foundation (NSF), under grants No.\ PHY--9972170 and No.\
PHY--0073034, respectively. The work of ADS is supported in part by
DOE Grant No.\ DE--FG01--00ER45832 and NSF Grant No.\ PHY--0071312.
\end{acknowledgments}


\end{document}